\documentclass[prb,twocolumn,showpacs]{revtex4}
\usepackage{graphicx}
\usepackage{amsmath}
\usepackage{amssymb}

\def\lsim{\
  \lower-1.2pt\vbox{\hbox{\rlap{$<$}\lower5pt\vbox{\hbox{$\sim$}}}}\ }
\def\gsim{\
  \lower-1.2pt\vbox{\hbox{\rlap{$>$}\lower5pt\vbox{\hbox{$\sim$}}}}\ }

\begin{document}
\title{On the probable wave nature of Bose crystals}
\author{M. D. Tomchenko}

\begin{abstract}
At the present time, it is considered that Bose crystals are formed at the cooling of a fluid, because the state
 of crystal is more favorable by energy. It is also believed \cite{rev1,rev3} that no ordering factor
 forming a crystal is present, except for the interatomic interaction.
 However, the available solutions \cite{rev1,rev3,nosanow1964} for the wave functions (WFs) of the ground and excited states
 of a crystal are approximate and are obtained for cyclic boundary conditions, which are not realized in the Nature.
Here, we present the exact solutions for the WFs of a Bose crystal with rectangular lattice under natural zero boundary conditions.
 The structure of WFs implies that 1) a crystal is formed by a standing  wave in the probability field;
2) a crystal in the ground state
 contains a condensate of atoms with the wave vector $\textbf{k}_{\textbf{l}}=(\pi/\bar{R}_{x}, \pi/\bar{R}_{y}, \pi/\bar{R}_{z})$ ($\bar{R}_{x}, \bar{R}_{y}, \bar{R}_{z}$
are the periods of the lattice) that is equal to a half of the vector of the reciprocal lattice.
These solutions indicate that the ordering factor forming a crystal is an intense standing wave similar to a sound one.
Thus, the periodicity of a lattice is caused by that of a sound wave,
but not only by the energy minimum principle.
Apparently, the crystals of other types and with different lattices have the wave nature as well.
The condensate opens a possibility to explain the nonclassical inertia moment discovered by Kim and Chan
\cite{kim-chan1,kim-chan2} in solid $He^4$,
which testifies, probably, to the presence of a superfluid subsystem in the crystal.
\end{abstract}

\pacs{61.50.Ah, 67.80.-s, 67.80.bd, 67.80.de}
\maketitle

The modern microscopic theory of Bose crystals is described in surveys \cite{rev1,rev3}.
Traditionally, the WF of the ground state (GS) of a crystal is sought in the form \cite{rev1,nosanow1964}
 \begin{equation}
   \Psi_{0} \approx  e^{-\sum\limits_{i>j}S_{BJ}(\textbf{r}_{i}-\textbf{r}_{j})}\prod\limits_{i=1}^{N} \varphi(\textbf{r}_{i}-\textbf{R}_{i}),
 \label{1-1}    \end{equation}
where $\textbf{r}_j$ and $\textbf{R}_j$ are the coordinates of atoms and lattice points,
$S_{BJ}$ is the Bijl-Jastrow function accounting for correlations, and $\varphi(\textbf{r})$ is written in the
approximation of small oscillations: $\varphi(\textbf{r})=e^{-\alpha^{2} r^{2}/2}$.

 WF (\ref{1-1}) is approximate. Moreover, a crystal lattice is postulated, and it is assumed that
the atoms perform small oscillations near lattice points.
The system of Bose particles described by such WF contains no condensate \cite{penronz1956}.

 We propose another, wave solution for $\Psi_{0}$ of a crystal. For a rectangular lattice, this solution takes the form
\begin{eqnarray}
 \Psi_{0}&=&e^{-\sum\limits_{i>j}S_{BJ}(\textbf{r}_{i}-\textbf{r}_{j}) +S_{w}^{(\textbf{l})}} \nonumber \\
 &\times&\prod\limits_{j=1}^{N}
 \{\sin{(k_{l_{x}}x_{j})}\sin{(k_{l_{y}}y_{j})} \sin{(k_{l_{z}}z_{j})}\},
  \label{1-2}    \end{eqnarray}
where $(k_{l_{x}}, k_{l_{y}}, k_{l_{z}})\equiv \textbf{k}_{\textbf{l}}=(l_{x}\pi/L_{x}, l_{y}\pi/L_{y}, l_{z}\pi/L_{z})$,
$l_{x}, l_{y}, l_{z}$ are integers, $L_{x}, L_{y}, L_{z}$ are sizes of the crystal
whose faces coincide with lattice planes, and
\begin{equation}
 S_{w}^{(\textbf{l})} \approx \sum\limits_{\textbf{q}=\textbf{q}^{res}}S_{1}^{(\textbf{l})}(\textbf{q})\rho_{-\textbf{q}}.
     \label{1-3} \end{equation}
Here, $\rho_{\textbf{q}} = \frac{1}{\sqrt{N}}\sum\limits_{j=1}^{N}e^{-i\textbf{q}\textbf{r}_j}$, à $\textbf{q}^{res}$ runs over the values $2\pi \left (\frac{j_{x}}{\bar{R}_{x}}, \frac{j_{y}}{\bar{R}_{y}},
 \frac{j_{z}}{\bar{R}_{z}} \right )$ with integers $j_{x}, j_{y}, j_{z}$, and $\bar{R}_{x}, \bar{R}_{y}, \bar{R}_{z}$
 are the periods of the lattice.

With the corrections (omitted due to their awkwardness) to the Bijl-Jastrow function and to $S_{w}^{(\textbf{l})},$ WF (\ref{1-2}) is the \textit{exact}
 solution of the Schr\"{o}dinger equation for $N$ interacting Bose particles in a box
 $L_{x}\times L_{y} \times L_{z}$ under zero boundary conditions. In fact,
WF (\ref{1-2})  is the exact solution at $l_{x}, l_{y}, l_{z}=1$, as is shown in \cite{zero-liquid}.
However, $l_{x}, l_{y}$ and $l_{z}$ are free integer parameters in the solution. So,
WF (\ref{1-2}) is the exact solution also at arbitrary integer $l_{x}, l_{y}, l_{z}\neq 0$.
 At $l_{x}=L_{x}/\bar{R}_{x}, l_{y}=L_{y}/\bar{R}_{y}, l_{z}=L_{z}/\bar{R}_{z},$
we have $\textbf{k}_{\textbf{l}}=(\pi/\bar{R}_{x}, \pi/\bar{R}_{y}, \pi/\bar{R}_{z}),$ and WF (\ref{1-2}) describes the GS of the crystal.
At $l_{x}, l_{y}, l_{z}=1,$ the same WF describes the GS of a Bose gas and a Bose liquid \cite{zero-liquid}.

The state of a crystal with a single longitudinal acoustic phonon is presented by WF
\begin{equation}
    \Psi_{\textbf{k}} =  (\psi^{0}_{\textbf{k}} + 7\,\mbox{permutations})\Psi_0,
  \label{1-4}     \end{equation}
  \begin{eqnarray}
&& \psi^{0}_{\textbf{k}} \approx  \rho_{-\textbf{k}} +
 \sum\limits_{\textbf{q} \neq 0, -\textbf{k}}
  Q_{1}(\textbf{q}, \textbf{k})\rho_{-\textbf{q}-\textbf{k}},
      \label{1-5}  \end{eqnarray}
where a permutation gives $\psi^{0}_{\textbf{k}}$ with the different sign of one or several components of the vector
$\textbf{k}$, and the components of the vector $\textbf{q}$ are quantized like $2\pi j/L$. WF (\ref{1-4}) describes a standing phonon.
It is possible to show that, by neglecting the ``coat'' of corrections, WF of the state with a single phonon
is equivalent to WF of the ground state (\ref{1-2}),
in which one of the wave vectors $\textbf{k}_{\textbf{l}}$ in the
bare product of sines is replaced by a wave vector of the form $\textbf{k}\pm \textbf{k}_{\textbf{l}}$.

WF (\ref{1-2}) indicates that the \textit{lattice of a crystal is formed by a standing wave
in the probability field; hence, the crystal has the wave nature}.
The antinodes of the wave correspond to points of lattice, and the period of the lattice is equal to a half-length of the assigning wave.
The previous paragraph implies that this standing wave is similar to $N$ longitudinal phonons.
Therefore, we may say that a crystal in the GS has $N$ standing
phonons with the wave vector $\textbf{k}_{\textbf{l}}$. In the excited states (with longitudinal sound),
a part of ``phonons'' of the GS is replaced by phonons with another wave vector.

We may expect that if the majority of assigning phonons of the GS are replaced by
phonons with another $\textbf{k}$, the crystal is melted.

The traditional solution (\ref{1-1}) was obtained without regard for boundaries or with the use of cyclic boundary conditions.
But the boundary conditions in the Nature are not cyclic and close to zero ones.
Thus, the explicit consideration of boundaries turns out to be essential and leads to the wave solution (\ref{1-2}).

The properties of solutions and their relationship to crystals are studied in \cite{zero-crystal} in detail.
Of the highest interest, in our opinion, are two following specific features.
1) It turns out that the atoms in the GS of a crystal do not merely oscillate around the equilibrium positions.
Rather, they perform the wave motion near some orbit. The orbit is the surface of a parallelepiped,
which is twice less than the crystal cell (which is also a parallelepiped). At the averaging over time,
the motion is similar to the oscillatory one. The higher the temperature of a crystal, the more chaotic the motion of atoms.
2) A crystal in the GS has a condensate of atoms with the wave vector $\textbf{k}_{\textbf{l}}$.
The condensate can be most simply revealed if the interatomic interaction is switched-off. Then WF (\ref{1-2}) is reduced to the
product of sines with the wave vector $\textbf{k}_{\textbf{l}}$: i.e., all $N$ atoms
of a crystal belong to the condensate with $\textbf{k}=\textbf{k}_{\textbf{l}}$. If the interaction is switched-on,
the number of atoms in the condensate decreases like that for a liquid.
In reality, the structure of a condensate is more complex. There are the condensates with $\textbf{k}=3\textbf{k}_{\textbf{l}}$ and, possibly,
with $5\textbf{k}_{\textbf{l}}$ (see \cite{zero-crystal}).

It follows from solution (\ref{1-2}) and the node theorem that the \textit{ground-state energy of a liquid
is always less than that of a crystal}.
Therefore, for all sorts of Bose atoms,
the state of underliquid with a lower energy than that of a crystal must exist (for more details, see \cite{zero-crystal}).
One well-known example of such a state is superfluid He II.

It is clear that the principle of formation of various crystals must be unique. We assume that it is a wave one.
To verify this assertion, it is necessary to find the solutions for natural Bose lattices such as the hcp, bcc, and fcc lattices
and charged Fermi lattices. For the last lattices, the wave solution is possible under the condition of the pairing of atoms and electrons,
at least on the level of correlations. To  our comprehension, the Bloch theorem is just related to such correlations.
The wave principle allows one to understand the nature of crystals in a new way. If a crystal is a wave,
then we can try to control its state with the help of external sound and electromagnetic waves.

The nonclassical inertia moment observed \cite{kim-chan1,kim-chan2} in solid
$He^4$  can be caused by the tunneling
of condensate atoms of the lattice through the crystal.
The condensate allows atoms to move synchronously, as a whole,
which separates a coherent subsystem in the crystal similarly to that in a liquid.
We note that a lot of experimental data on the supersolid phenomenon are available, and the general picture seems quite complicated.
In \cite{zero-crystal}, it is shown that the idea of
the tunneling of condensate atoms can be a key for the disentangling of this puzzle.

       \end{document}